\documentclass[twocolumn,aps,amssymb,prl]{revtex4}
\usepackage{graphicx}
\usepackage{epsfig,amsmath}

\begin{document}

\title{Shot noise of a multiwalled carbon nanotube field effect transistor}

\author{F.~Wu$^1$, T.~Tsuneta$^{1}$, R.~Tarkiainen$^{1}$,
D.~Gunnarsson$^{1}$, T.~H.~Wang$^{2}$, and P.~J.~Hakonen$^1$}
\affiliation{$^1$Low~Temperature~Laboratory,~Helsinki~University~of~Technology,~Finland
\\$^2$Institute of Physics, Chinese Academy of Sciences, Beijing, China}

\date{\today} 

\begin{abstract}
We have investigated shot noise in a 6-nm-diameter, semiconducting
multiwalled carbon nanotube FET at 4.2 K over the frequency range
600 - 950 MHz. We find a transconductance of 3 - 3.5 $\mu$S for
optimal positive and negative source-drain voltages $V$. For the
gate referred input voltage noise, we obtain 0.2 and 0.3 $
\mu\textrm{V}/ \sqrt{\textrm{Hz}}$ for $V>0$ and $V<0$,
respectively. As effective charge noise this corresponds to $2-3
\cdot 10^{-5}$ e/$\sqrt{\textrm{Hz}}$.

\end{abstract}
 \bigskip

\maketitle


Semiconducting single-walled carbon nanotubes have been shown to
provide extraordinary field effect transistors (FET)
\cite{dekker,dainature} in which the modulation of Schottky barriers
is often an important factor \cite{Heinze02,Appenzeller02}.
Intrinsic performance limits of these devices due to the mobility of
charge carriers have been investigated recently
\cite{Javey04,extramobility,contactSWNTFET,mceuen05,ephballistic,fuhrer05}.
Transconductances up to $g_m=\frac{\Delta I_{ds}}{\Delta V_g}=8700
\mu \textrm{S}/\mu \textrm{m}$, relating the change in drain-source
current $I_{ds}$ to gate voltage $V_{g}$, have been reported in
SWNTs on top of a high-$\kappa$ material (SrTiO$_3$)
\cite{fuhrer04}. It has been shown experimentally that $g_m$
increases as $\phi^2$ with the tube diameter $\phi$ \cite{mceuen05}.
This, however, takes place at the expense of a reduced energy gap,
which sets an upper limit for the diameter of room temperature
devices.

Another important issue for typical FET applications is the noise
power generated by the device. Here we are interested in the
uncoupled noise performance, the understanding of which is a
prerequisite for the proper noise minimization with a finite
 source impedance. In general, the low-frequency current noise $S(\omega) = \int
e^{i\omega t} \langle \delta i(t) \delta i(0) \rangle $ in a
mesoscopic sample can be written as

\begin{equation}\label{FullNoise}
    S= \frac{4 k_B T}{R}(1-F) + F 2eI \coth \left(\frac{eV}{2k_B T}\right)
\end{equation}
where $R$ is the resistance of the sample, \textit{T} is
temperature, \textit{F} denotes the Fano-factor, and \textit{V}
 is the DC biasing voltage. The Fano-factor depends on
 transmission coefficients of the transport
 channels of the sample, as well as on inelastic processes causing energy
 relaxation, which are known to lower the shot noise
 \cite{BB}. The best uncoupled performance  corresponds to the minimization of
 $S/g_m^2$ which yields the minimum equivalent voltage noise at the
 input. Here we present the first experimental determination of this
 noise quantity in a semiconducting nanotube device.


In our 4-K measurement setup, IV characteristics and differential
conductance properties are measured in a regular two terminal
configuration, supplemented with a radio-frequency noise
amplification circuitry. Bias-tees are used to separate dc bias and
the current-dependent noise signal at radio frequencies. We use a
low-noise, cooled amplifier \cite{Cryogenics04} with working
frequency range of 600 - 950 MHz for suppressed 1/\textit{f} noise.
The total gain of the amplifier chain amounts to 80 dB (16 dB at 4.2
K) and the noise temperature of the whole setup is roughly 10 K; for
detection, we used a zero-bias Schottky diode. A switch and a
high-impedance tunnel junction are used to calibrate the gain and
the bandwidth, \textit{i.e.}, we can determine the Fano-factor of
our CNT samples by direct comparison with the noise measured on a
tunnel junction sample having $F=1$.

We determine the Fano-factor at drain-source voltage $V_{ds}$ as

\begin{equation}\label{FanoExp}
    F = \frac{S(I_{ds}) - S(0)}{2eI}=\frac{1}{2eI_{ds}}\int_{0}^{I}
    \left(\frac{dS}{dI_{ds}}\right) dI
\end{equation}
where $\left(\frac{dS}{dI_{ds}}\right)$ represents the
differentially measured noise using a small modulation voltage of
0.5 mV at 18.5 Hz on top of $V_{ds}$. At large currents, this
determination coincides with the ordinary definition of Fano-factor.
In the intermediate bias region, there will be corrections that
depend on the ratio of differential resistance
$\frac{dV_{ds}}{dI_{ds}}$ to $V_{ds}/I_{ds}$ due to thermal noise
coupling, but these corrections are negligible for the analysis in
this paper \cite{NOTE,details}. Because the sample impedance is not
matched to the preamplifier, we are able to measure shot noise only
at currents of $I > I_{th}$ where  $ F I_{th}$ must be around 0.01
$\mu$A.

Our tube material, provided by the group of S. Iijima, was grown
using plasma enhanced growth without any metal catalyst
\cite{Koshio}. The tubes were dispersed in dichloroethane and, after
15 min of sonication, they were deposited on to thermally oxidized,
strongly doped Si wafers. A tube of 4$-\mu$m in length was located
with respect to alignment markers using a FE-SEM Zeiss Supra 40.
Subsequently, Ti contacts of width 900 nm were made using standard
overlay lithography: 10-nm titanium layer was covered by 70 nm Al in
order to facilitate proximity induced superconductivity at subkelvin
temperatures. The length of the tube section between the contacts
was 1200 nm. The electrically conducting body of the silicon
substrate was employed as a back gate, separated from the sample by
100 nm of SiO$_2$. The sample was bonded to a sample holder with
miniature, 6-GHz bias tees using 25 $\mu$m Al bond wires with less
than 10 nH of inductance.

\begin{figure}

    \includegraphics[width=7cm]{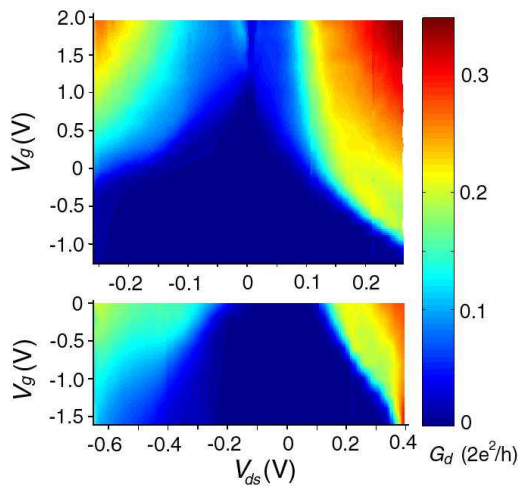}

    \caption{(color on line) Normalized differential conductance $G_d/G_0$ with $G_0=2e^2/h$
    for our semiconducting sample measured at 4.2 K on the gate $V_g$ vs. bias voltage $V_{ds}$ plane:
    the color scale is given by the bar on the right.
    For the sample parameters, see text.} \label{IV}

    \end{figure}

Differential conductance  $G_{d}=\frac{dI_{ds}}{dV_{ds}}$ for our
sample is illustrated in Fig. \ref{IV} in units of $G_0=2e^2/h$.
$G_{d}$ is seen to display a roughly linear conductance, on the
order of 0.1 $G_0$, at voltages $V_{ds}=-0.1...+0.1$ V and
$V_g=1-4$ V. When current is increased to $I_{ds}=1$ $\mu$A,
$G_{d}$ becomes on the order of 0.2 $G_0$, which is a typical
value for metallic PECVD tubes of the same batch. Thus, there is
no obvious difference in conductance between semiconducting and
metallic specimens as observed in SWNT tubes \cite{Mceuen99}.

Our nanotube sample is clearly n-type doped initially. Looking at
$G_d(V_{ds},V_g)$, we deduce that the charge neutrality point
 is located around $V_g=-1.7$V which also corresponding to the maximum gap of 0.8
 V.
Using the gate capacitance $C_g=18$ aF and total island capacitance
of 0.4 fF (see below), we find that the initial shift of the
Fermi-level from the charge neutrality point is approximatively
-0.25 V. This differs substantially from the value of +0.4 V that
has been reported for MWNTs \cite{doping}. Typically, n-type doping
in NTs has been obtained only using potassium deposition
\cite{Bockrath00,Appenzeller04}.

The capacitance of our backgate was measured by observing Coulomb
modulation in the range $V_g= 2 ... 4$ V. The measured periodicity
of 8.8 mV corresponds to 18 aF. The island capacitance
$C_{\Sigma}=0.4$ fF was estimated using a geometric capacitance of
$C=200$ pF/m in series with a quantum capacitance of similar
magnitude along the full length 4$\mu$m. Owing to local thinning of
the 100-nm SiO$_2$ oxide due to Al wire bonding, we used gate
voltages only up to $\pm 4$ V in our studies. In addition, we
limited our measurements for currents below 5 $\mu$A which is on the
same order as typical ON-state current in SWNT devices.


The model, that we employ to account for our basic findings, has
been proposed and discussed in Ref. \cite{Park01}. There it was
conjectured that owing to band pinning at the metal-nanotube
interfaces, small quantum dots are formed at the ends of the
nanotube when the tube is strongly doped by the gate voltage. In our
case, this is corroborated by the appearance of another
(quasiperiodic) gate modulation in the range $V_g=-0.6 ... -4$ V.
This gate period changed from $\Delta V_g=0.13$ V at $V_g \sim -1$ V
to $\Delta V_g=0.18V$ at $V_g \sim -2.5$. The size of the period is
in accordance with the findings in Ref. \onlinecite{Park01}, while
the increase in $\Delta V_g$ in our data reflects a decrease in the
dot size as $V_g$ becomes more strongly negative.

The expected gap for a semiconducting tube of diameter $\phi =6$ nm
is approximatively $V_{gap}=0.14$ V \cite{Saito}. If the extra width
of the gap were due to the quantum dots at the ends of the tube,
their capacitance would be about 1 aF, i.e $<10$ nm in length. Room
temperature measurements indicate that the gap indeed is composed of
a few smaller components, but we cannot exclude the possibility
that, at some large gap value, the tube is broken into more than
three quantum dots.

In CNT-FETs, the signature of charge carrier freeze-out in $I_{ds}$
vs. $V_g$ sweeps is the appearance of a threshold voltage, related
to current by the form $I_{ds}^2 \propto V_g-V_{th}$
\cite{fuhrer04}. Using this form, we obtain $V_{th}=0.25V$ for the
pinch-off. When lowering the gate voltage towards $V_{th}$, the
small voltage IV curves change from linear to more and more
power-law-like: in the range $V_g = 1 \textrm{V}... V_{th}$, the
exponent varies from 1 to 3 (in Fig. \ref{IV}, the exponent of $G_d$
varies from 0 to 2, respectively).

   \begin{figure}

    \includegraphics[width=7cm]{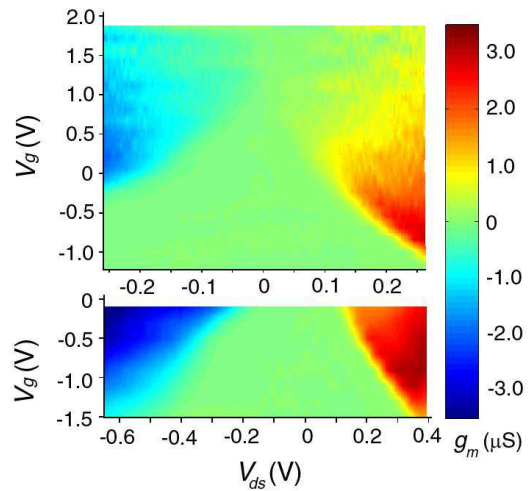}

    \caption{Transconductance $ g_m$ as a function of bias $V_{ds}$ and gate voltage $V_g$.
    } \label{Transconductance}

    \end{figure}

Measured transconductance around the pinch-off region is displayed
in Fig. \ref{Transconductance} The largest magnitude of
transconductance is roughly equal at positive and negative bias:
$\sim3$ $\mu$S at $V > 0$ and $\sim3.5$ $\mu$S at $V < 0$. At
positive bias the optimum is reached in a small region of bias
values around $V_{ds}=0.37$ V and $V_{g}=-0.9$ V  whereas at $V < 0$
the maximum value is obtained on a more extended region at
$V_{ds}<-0.5$ V around $V_{g}=0$.


\begin{figure}

    \includegraphics[width=8cm]{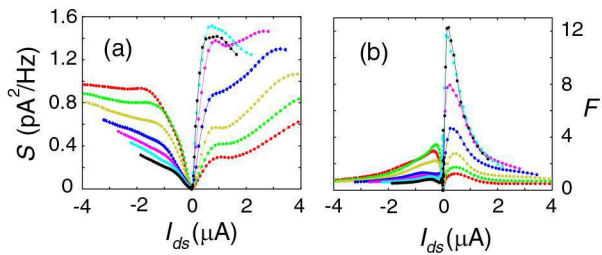}

    \caption{Current noise $S$ integrated over the frequency range 600 - 950
    MHz. vs.  current $I_{ds}$ (a), and the
    corresponding Fano-factor (b). Due to lack of sensitivity, currents below
    0.01 $\mu$A have been cut off from
    the plot. The bias voltage varies over $V_{ds}=-1.2$ ... 0 V
    in steps of 0.2 V (from top to bottom at $V_{ds}>0$ and from bottom to top at
    $V_{ds}<0$)
    } \label{NoiseCurves}

    \end{figure}

Fig. \ref{NoiseCurves}a illustrates the measured current noise in
the range $V_g = -1.2 ... -0$ V which is right below the pinch-off
of threshold $V_{th}$; the corresponding Fano-factor is given in
Fig. \ref{NoiseCurves}b. At large negative bias, and with large
positive bias at $V_g << V_{th}$, the noise can be regarded as shot
noise from an asymmetric double junction system
\cite{korotkov,Beenakker1996PhysicaA}, which yields $F= (\Gamma_1^2+
\Gamma_2^2)/(\Gamma_1 +\Gamma_2)^2 < 1$ where $\Gamma_1$ and
$\Gamma_2$ refer to tunneling rates in the two tunnel barriers. At
small $V_{ds}$, especially at $V_{ds} >0$, the measured noise is
strongly peaked, and the corresponding Fano-factor reaches $F=12$ at
its maximum. This behavior may be an indication of noise due to
inelastic co-tunneling as argued by Kouwenhoven and coworkers in a
SWNT quantum dot at small bias \cite{Kouwenhoven06}. In our case,
however, we believe that a more likely explanation is due to a
bias-dependent fluctuator that modulates the transmission at one of
the contacts \cite{ReetaPhysicaE05}. According to this model, the
peak in the noise vs. current reflects the movement of the corner
frequency of the Lorentzian fluctuation spectrum across the
frequency band of the measurement. Initially, the noise increases
when the corner frequency approaches the measurement band from
below. The decrease at large currents is because the total
integrated noise over the Lorentzian spectrum is fixed, and as the
corner frequency continues to grow, the noise per unit band has to
decrease \cite{ReetaPhysicaE05}. Thus, we argue that there are
bias-dependent fluctuators in metal-nanotube systems with tunneling
rates in the GHz regime.

\begin{figure}

    \includegraphics[width=7cm]{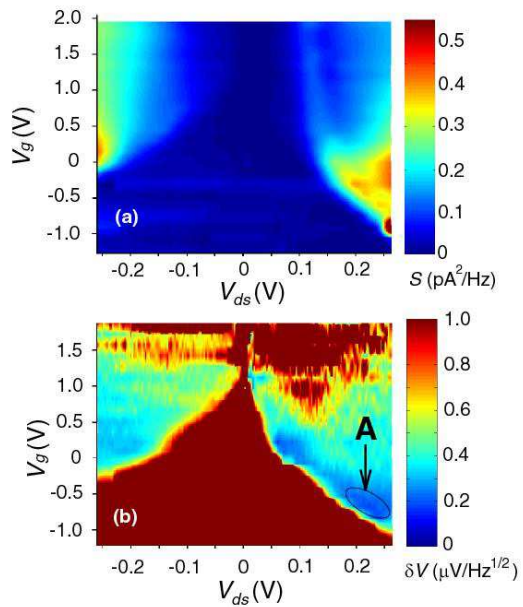}

    \caption{a) Current noise  $S$ over $V_g$ \emph{vs.} $V_{ds}$ plane.
    b) Noise power of (a) converted into input voltage
    noise by dividing by $ g_m^2$. The region of smallest noise
    has been denoted by an ellipsoid.
    } \label{Noise}

    \end{figure}

The overall noise characteristics of our device are illustrated in
Fig. \ref{Noise}a. By multiplying $S$ with $1/g_m^2$, we may convert
the measured current noise into voltage noise at the gate, which is
displayed in Fig. \ref{Noise}b. Here we assume that the electrical
properties of the tube do not change with frequency up to 800 MHz,
as indicated by the experiments by Burke and coworkers \cite{Yu05}.
The lowest-noise region of operation is marked by A, which is very
close to the region of maximum $g_m$. However, since the variation
of $g_m$ is rather slow with $V_{ds}$ and $V_g$, the optimum noise
is located at a local minimum of noise power. Note that, even though
negative bias provides larger $g_m$ and  smaller Fano-factors, the
smallest input equivalent voltage noise $\delta V_g$ is found at
$V>0$, because $I_{ds}$ is much smaller at optimum regions at $V>0$
than at $V<0$.

At point A, we find $\delta V_g = 0.2 \mu
\textrm{V}/\sqrt{\textrm{Hz}}$. This input voltage noise, in turn,
can be converted into charge noise at the gate, which yields $\delta
q_g = 20$ $ \mu \textrm{e}/\sqrt{\textrm{Hz}}$. At negative bias,
our results are about 30 \% worse, \textit{i.e.} $\delta q_g = 30 $
$\mu \textrm{e}/\sqrt{\textrm{Hz}}$ These values are close to the
results obtained in RF-SET setups using impedance matching
\cite{JAP}. Note that no matching circuits have been employed here
and that the noise has been measured over a large band of 600 - 950
MHz. In a RF-SET setup with hundreds of parallel SET's, a bandwidth
of 1 GHz has been achieved, but with a limited charge sensitivity of
$\delta q_g = 2 $ $\textrm{me}/\sqrt{\textrm{Hz}}$ due to a large
input capacitance \cite{Gustavsson06}. Thus, our results suggest
that nanotube FETs based on MWNTs may be employed as sensitive
charge detectors at high frequencies, rivaling the performance of
RF-SETs.

\begin{figure}

    \includegraphics[width=7cm]{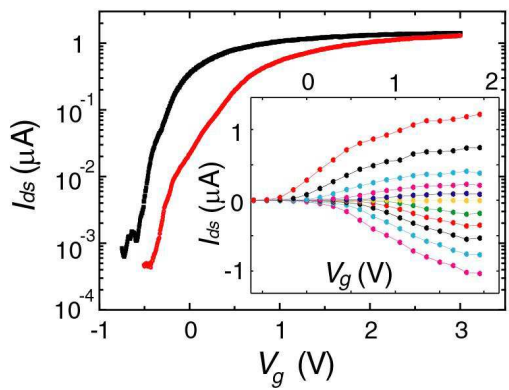}

    \caption{Nanotube current as a function of gate $V_g$ bias voltage
    $V_{ds}=+0.135$ V ($\square$) and $V_{ds}=-0.135$ V ($\circ$). The
    inset displays a set of current traces on linear scale measured when at $V_{ds}$
    has been stepped from $- 0.13$ V to 0.13 V by 26 mV (from bottom to top).
    } \label{gatesweep}

 \end{figure}

Semiconducting nanotube devices are often described in terms of
field-effect (FE) mobility $\mu_{FE}=\frac{L}{C^*_g}
\frac{\partial G}{\partial V_g} = \frac{L}{C^*_g
V_{ds}}\frac{\partial I}{\partial V_g}$ where $C^*_g$ denotes gate
capacitance per unit length \cite{fuhrer04}. This quantity is
employed for the description of the "bulk" properties of the tube
when the contribution from the contacts can be neglected. In our
case, even though the length of the tube is not extremely large,
the resistance of the tube should dominate close to the pinch-off
of the device. Gate sweeps at $V_{ds} = \pm0.135$ V are
illustrated in Fig. \ref{gatesweep}. From the figure we may read
at 0.1 $\mu$A that conductance $I_{ds}/V_{ds}$ changes by
decade/250mV and by decade/500mV at positive and negative bias,
respectively. This fact that the threshold is sharper at $V_{ds} >
0$ is visible also from the inset of Fig. \ref{gatesweep} which
displays a set of current traces vs. $V_g$ measured with stepping
$V_{ds}$ over $- 0.13$ V to 0.13 V. These data at 4.2 K yield for
the maximum FE-mobility $\mu_{FE}=1$ m$^2$/Vs which falls short by
a factor of $>100$ from the value extrapolated for $\phi=6$ nm
using temperature-scaled data of Ref. \onlinecite{mceuen05}
measured above 50 K for SWNTs with $\phi= 1-4$ nm. This
discrepancy indicates that our MWNTs are more strongly diffusive
than typical semiconducting SWNTs.

In summary, we have presented first noise investigations on
semiconducting nanotube FETs. We find noise behavior that varies
between sub- and super-Poissonian values. The sub-Poissonian
values are consistent with double Schottky barrier configuration
while the super-Poissonian results indicate the presence of two
level fluctuators with bias-dependent switching rates exceeding 1
GHz. For the input referred noise, expressed in terms of charge
noise on the gate, we find $2-3 \cdot 10^{-5}$
e/$\sqrt{\textrm{Hz}}$. Thus, these devices may challenge regular
aluminum-based RF-SETs as the ultimate charge detectors.

We thank S. Iijima, A. Koshio, and M. Yudasaka for the carbon
nanotube material employed in our work. We wish to acknowledge
fruitful discussions with L.~Lechner, M.~Paalanen, B. Placais, and
L.~Roschier. This work was supported by the TULE programme of the
Academy of Finland and by the EU contract FP6-IST-021285-2.


\begin{thebibliography}{99}
\bibitem{NOTE} At liquid helium temperature, the crossover
between thermal and shot noise takes place at $V_{ds} \sim 1$ mV,
which is a small voltage compared with typical $V_{ds}$ values
employed in our measurements.
\bibitem{ReetaPhysicaE05} R. Tarkiainen, L. Roschier, M. Ahlskog, M.
Paalanen, and P. Hakonen, Physica E \textbf{28}, 57 (2005). Note
that the corner frequency of the fluctuator may go down or up with
the bias and the result is qualitatively the same.
\end{thebibliography}
\end{document}